\documentclass[12pt,a4paper]{article}
\usepackage{amsmath,amsfonts,amssymb,graphicx}
\usepackage[colorlinks]{hyperref}

\begin{document}

\title{Leptonic asymmetry  of the sterile neutrino hadronic decays in the $\nu MSM$ }

\author{Volodymyr M. Gorkavenko\thanks{E-mail:
gorka@univ.kiev.ua}, Igor Rudenok\thanks{E-mail:igrudenok@gmail.com},\\ and  Stanislav I. Vilchynskiy\thanks{E-mail: sivil@univ.kiev.ua}\\
\it \small Department of Physics, Taras Shevchenko National
University of Kyiv,\\\small\it 64 Volodymyrs'ka St., Kyiv, 01601,
Ukraine} \date{}

\maketitle

\begin{abstract}
We consider  the leptonic asymmetry  generation in the $\nu MSM$ via
hadronic decays of  sterile neutrinos at $T\ll T_{EW}$, when the
masses of  two heavier sterile neutrinos are between $m_\pi$ and 2
GeV.  The choice of upper mass bound  is motivated by absence of
direct experimental searches for singlet fermions with greater mass.
We carried out computations at zero temperature and ignored the
background effects. Combining constraints of sufficient value of the
leptonic asymmetry for production of dark matter particles,
condition for sterile neutrino to be out of thermal equilibrium and
existing experimental data we conclude that it can be satisfied only
for mass of heavier sterile neutrino in the range $1.4$ GeV
$\lesssim M <2$ Gev and only for the case of normal hierarchy for
active neutrino mass.

\end{abstract}

\section{Introduction}
\label{intro}

The Standard Model (SM)  is minimal relativistic field theory, which
is able to explain almost all particle physics experimental data
\cite{SM}. However, there are several observable facts, that cannot
be explained in the SM frame.  Firstly, the neutrinos of SM are
strictly massless, that contradict to the experimental fact of the
neutrinos oscillations \cite{PG,Strumia}. The second problem is the
impossibility to explain the baryon asymmetry of the Universe (BAU)
within the SM. Finally, the SM does not provide the dark matter (DM)
candidate. Also the SM can not solve the strong CP problem in
particle physics, the primordial perturbations problem and the
horizon problem in cosmology, etc.

The solutions of the  above mentioned problems of the SM require
some new physics between the electroweak and the Planck scales. An
important challenge for the theoretical physics is to see if it is
possible to solve them using only the extensions of the SM below the
electroweak scale \cite{last}.

The Neutrino Minimal Standard Model ($\nu MSM$) is an extension of
the SM by three massive right-handed neutrinos (sterile neutrinos),
which do not take part in the gauge inter\-actions of the SM
\footnote{This is why these neutrinos are called sterile neutrinos.
The left-handed neutrinos of the SM are called active neutrinos.}.
The model was suggested by M. Shaposhnikov and T. Asaka
\cite{Shap1,Shap2}. The masses of sterile neutrinos are predicted to
be smaller than electroweak scale, and thus there is no new energy
scale introduced in the theory. The parameters of the $\nu MSM$ can
be chosen in order to explain simultaneously the masses of active
neutrinos, the nature of DM, and BAU.

 The lightest sterile neutrino (the mass is expected to be
in the KeV range \cite{last}) can be intensively produced in the
early Universe and have cosmologically long life-time.  So, it might
be a viable DM candidate. The sufficient amount of this neutrinos
can be generated through an efficient resonant mechanism proposed by
Shi and Fuller \cite{Shi}.

In the $\nu MSM$ the required amount of the leptonic asymmetry (in
accordance with Shi and Fuller mechanism) can be created due to
decays of the two heavier sterile neutrinos. This particles are
generated at temperature $T> T_{EW}$ and their masses are expected
to be in range $m_\pi<M_I<T_{EW}$ \cite{Sh3}, where $m_\pi$ is the
pion mass
 and $M_I$ is the mass of $I$-sterile neutrino. The
leptonic asymmetry at the temperature of the sphaleron freeze-out
($T\sim T_{EW}$) is related to the baryon asymmetry of the Universe.
At temperature $T< T_{EW}$ the leptonic asymmetry from decays of
heavier sterile neutrinos can not convert into the baryon asymmetry
and is accumulated. As it was shown in \cite{last,Sh1} the required
amount of the leptonic asymmetry $\Delta=\Delta L/L=(n_L-n_{\bar
L})/(n_L+n_{\bar L})$:
\begin{equation}\label{asymmetry}
10^{-3}<\Delta<2/11
\end{equation}
 has to already exist in the Universe at the moment of
the beginning of production of the DM particles (takes place at the
temperature around 0.1 GeV).

We consider here the leptonic asymmetry  generation at $T\ll
T_{EW}$, when the masses of  two heavier sterile neutrinos are
between $m_\pi$ and 2 GeV. The motivation is following. The mass of
heavier sterile neutrino can not be less then $m_\pi$ (the
constraint is coming from accelerator experiments combined with Big
Bang Nucleosynthesis (BBN) bounds \cite{Gorbunov,Kusenko}) and there
is no direct experimental searches for singlet fermions with mass
more then 2 Gev \cite{Gorbunov}.

Since the masses of active neutrinos in the $\nu MSM$ are produced
by the "see-saw"\, mechanism \cite{2} some constraints on the
parameters of the $\nu MSM$ come from active neutrino parameters
that can be found from the experiments on the neutrino oscillations.
Namely, these are the mass squared differences of active neutrinos
and the mixing angles. Until recently the mixing angle $\theta_{13}$
was supposed to have a close to zero value. But  new observations
indicated its essential difference from zero \cite{T2Ktheta13}.

The aim of this work is to obtain constraints on the parameters of
the $\nu MSM$ from the required amount of the leptonic asymmetry and
cosmology conditions. Also we want to investigate  the influence of
non-zero mixing angle $\theta_{13}$ on space of the allowed
parameters of the $\nu MSM$. We do it following \cite{Tibor} using a
simple model: we ignore the background effects and do computations
at zero temperature.

The paper is organized as follows. In Section \ref{rozdil2} we
present the Lagrangian of the $\nu MSM$, make its convenient
parametrization and present the Yukawa couplings in terms of active
neutrinos mass matrix parameters. In Section \ref{rozdil3} we derive
the expression for the leptonic asymmetry.
 The  limitations on the $\nu MSM$ parameters are imposed in Section
\ref{rozdil4}. Section \ref{rozdil5} is devoted to the analysis and
conclusions.

\section{Basic formalism of the $\nu MSM$}
\label{rozdil2}

In the $\nu MSM$  \cite{Shap1,Shap2} the following terms  are added
to the Lagrangian of the SM (without taking into account the kinetic
terms):
\begin{equation}\label{lagdiag0}
\mathcal L^{ad}=-F_{\alpha I}\bar L_\alpha \tilde\Phi
\nu_{IR}-\frac{M_{IJ}}{2}\bar \nu_{IR}^c\nu_{JR}+h.c.,
\end{equation}
where index  $\alpha=e,\mu,\tau$  corresponds to the active neutrino
flavors, indices  $I,J$ run from $1$ to $3$, $L_{\alpha}$ is for the
lepton doublet of the left-handed particles, $\nu_{IR}$ is for the
field functions of the sterile right-handed neutrinos, the
superscript $\raisebox{-0.8em}{$"$}\!c"$ means charge conjugation,
$F_{\alpha I}$ is for the new (neutrino) matrix of the Yukawa
constants, $M_{IJ}$ is for the Majorana mass matrix of the
right-handed neutrinos, $\Phi$ is for the field of the Higgs
doublet, ${\tilde\Phi}=i\sigma_2\Phi^*$.


After the spontaneous symmetry breaking the field of the Higgs
doublet in unitary gauge is
$$\Phi=\left(\begin{array}{c}0\\ \frac{v+h}{\sqrt{2}}\end{array}\right),$$
where $h$ is the neutral Higgs field and the parameter $v$
determines minimum of the Higgs  field potential
 $(v\cong247\,\,\mbox{GeV})$. In this case  Lagrangian
\eqref{lagdiag0} acquires the Dirac-Majorana neutrino mass terms:
\begin{equation} \label{2}
\mathcal L^{DM}=-\frac{v}{\sqrt{2}}F_{\alpha
I}\bar{\nu}_\alpha\nu_{IR}-\frac{M_{IJ}}{2}\bar
\nu_{IR}^c\nu_{JR}+h.c.,
\end{equation}
or in conventional form \cite{Belenki}
\begin{equation}\label{dm1}
\mathcal L^{DM}=-\left(\overline{(N_L)^c}
\,\frac{M^{DM}}2\,N_L+h.c.\right),
\end{equation}
where
\begin{equation}\label{dm2}
N_L=\left(%
\begin{array}{c}
  \nu_L \\
  \nu_R^c \\
\end{array}
\right)\!;\,\, N_L^c=\left(
\begin{array}{c}
  \nu_L^c \\
  \nu_R \\
\end{array}
\right)\!;\,\,
M^{DM}=\left(%
\begin{array}{cc}
  M_L&M_D{}^T \\
  M_D&M_R \\
\end{array} \right)\!
\end{equation}
and
\begin{equation}\label{dm1a} M_L=0,\quad
M_D=F^+\frac{v}{\sqrt2}\,,\quad M_R=M^*,
\end{equation}
where $M,F$ are square matrix of the third order with elements
$F_{\alpha I}$ and $M_{IJ}$.

In  zero approximation the $\nu MSM$ Lagrangian is assumed to be
invariant under $U(1)_e\times U(1)_\mu\times U(1)_\tau$
transformations, that provides preservation of the $e, \mu, \tau$
lepton numbers separately. It is also assumed that two heavier
sterile neutrinos interact with the active neutrinos, but the third
(lightest) sterile neutrino does not interact\footnote{Therefore the
lightest sterile neutrino in the $\nu MSM$ is a candidate for the DM
particle.}. This
 assumption can be realized by following matrix $M^{DM}$\cite{sb}:
\begin{equation}
M_R^{(0)}=\left(\begin{array}{ccc}0&0&0\\0&0&M\\0&M&0\end{array}\right),\qquad
M_D^{(0)+}=\frac{v}{\sqrt{2}}\left(\begin{array}{ccc}0&h_{12}&0\\0&h_{22}&0\\0&h_{32}&0\end{array}\right),\qquad
M_L^{(0)}=0
\end{equation}

In this approximation  we have two massive sterile neutrinos with
equal mass $M$, the third neutrino is massless, and all active
neutrinos have zero mass. It contradicts observable data
\cite{PG,Strumia}. To adjust it next small  terms are added to the
matrix $M^{DM}$ \cite{sb}:
\begin{multline}
M_R^{(1)}=\Delta
M=\left(\begin{array}{ccc}m_{11}e^{-i\alpha}&m_{12}&m_{13}\\m_{12}&m_{22}e^{-i\beta}&0\\m_{13}&0&m_{33}e^{-i\gamma}\end{array}\right),\\
M_D^{(1)+}=\frac{v}{\sqrt{2}}\left(\begin{array}{ccc}h_{11}&0&h_{13}\\h_{21}&0&h_{23}\\h_{31}&0&h_{33}\end{array}\right),
M_L^{(1)}=0
\end{multline}

This correction violates $U(1)_e\times U(1)_\mu\times U(1)_\tau$
symmetry, leads to the appearance of the mass of the third sterile
neutrino and takes off the mass degeneracy for two heavier sterile
neutrinos. It's also leads to the appearance of the extra small
masses of the active  neutrinos  and nonzero mixing angles among
them.

In the terms of the introduced  corrections Lagrangian
\eqref{lagdiag0} is
\begin{equation}\label{nongiag1}
\mathcal{L}^{ad}=-h_{\alpha
I}\bar{L}_\alpha\tilde{N_I}\tilde{\Phi}-M\bar{\tilde{N_2^c}}\tilde{N_3}-\frac{\Delta
M_{IJ}}{2}\bar{\tilde{N_I^c}}\tilde{N_J}+h.c.,
\end{equation}
where $\tilde{N_I}$ are right-handed neutrinos in the gauge basis.

In order to find the masses of the active neutrino one has to make
the diagonalization of the matrix $M^{DM}$. The diagonalization
undergoes in two steps. Firstly, $M^{DM}$ matrix is reduced to the
block-diagonal form via the unitary transformation \cite{see-saw} in
the "see-saw"\, approach:
\begin{equation}
M_{block}=W^TM^{DM}W=\left(\begin{array}{cc}-(M_D)^T (M_R)^{-1}
M_D&0\\0&M_R\end{array}\right)=\left(\begin{array}{cc}M_{light}&0\\0&M_{heavy}\end{array}\right),
\end{equation}
where
\begin{equation}\label{ss3}
W=\left(
\begin{array}{cc}
  1-\frac{1}{2}\varepsilon^+\varepsilon & \varepsilon^+ \\
  -\varepsilon & 1-\frac{1}{2}\varepsilon\varepsilon^+ \\
\end{array}%
\right),\quad \varepsilon=M_R^{-1}M_D\ll1
\end{equation}
and $M_{light}=-(M_D)^T (M_R)^{-1} M_D$, $M_{heavy}=M_R$ are  the
mass matrix of the active and sterile neutrinos respectively. Now
each block of the matrix $M^{DM}$ may be diagonalized independently
by the matrix
\begin{equation}\label{ss4a}
U=\left(%
\begin{array}{cc}
  U_1 & 0 \\
  0 & U_2 \\
\end{array}%
\right).
\end{equation}

The mass matrix of the active and sterile neutrinos is diagonalized
by unitary trans\-for\-mation $U_{1(2)}$:
\begin{equation}
U_1^TM_{light}U_1=diag(m_1,m_2,m_3),\quad
U_2^TM_{heavy}U_2=diag(M_1,M_2,M_3).
\end{equation}

There is a standard parametrization
 \cite{PG} for $U_{1(2)}$:
\begin{multline}\label{ss8}
U_{1(2)}=\left(\!\!
\begin{array}{ccc}
  c_{12}c_{13}  & c_{13}s_{12} &  s_{13}e^{-i \delta}  \\
    -s_{12}c_{23}-c_{12}s_{23}s_{13}e^{i \delta}  & c_{12}c_{23}-s_{12}s_{23}s_{13}e^{i \delta} & s_{23}c_{13}  \\
  s_{12}s_{23}-c_{12}c_{23}s_{13}e^{i \delta} & -c_{12}s_{23}-s_{12}c_{23}s_{13}e^{i \delta} & c_{23}c_{13}  \\
\end{array}%
\!\!\right)\times\\\times\left(\!\!
\begin{array}{ccc}
  e^{i \alpha_1 /2} & 0 & 0 \\
  0 & e^{i \alpha_2 /2} & 0 \\
  0 & 0 & 1 \\
\end{array}%
\!\!\right)\!\!,
\end{multline}
where $c_{ij}=\cos \theta_{ij}$, $s_{ij}=\sin \theta_{ij}$,
$\theta_{12},\theta_{13},\theta_{23}$ are the three mixing angles;
$\delta$ is the Dirac phase, and $\alpha_1, \alpha_2$ are the
Majorana phases. The angles  $\theta_{ij}$ can be  in the region
$0\leq\theta_{ij}\leq\pi/2$, phases $\delta,\alpha_1, \alpha_2$ vary
from  $0$ to $2\pi$. Each of the matrices $U_{1}$ and $U_{2}$
contains its own, independent angles and phases.

Then the elements of the $M_{light}$  can be defined by masses and
elements of mixing matrix $U$ of the active neutrinos:
\begin{equation}
[M_{light}]_{\alpha\beta}=m_1U^*_{\alpha1}U^*_{\beta1}+m_2U^*_{\alpha2}U^*_{\beta2}+m_3U^*_{\alpha3}U^*_{\beta3}.
\end{equation}
The data that come from the neutrino oscillation experiments are
presented in Tab.1:

\begin{center}
\begin{tabular}{c c}\hline
 central value & 99$\%$ confidence interval \\
\hline  $\Delta m^2_{21} = (7.58\pm0.21)\cdot10^{-5}\,eV^2$&
 $(7.1 - 8.1)\cdot10^{-5} eV^2$\\
 $|\Delta m^2_{23}| = (2.40 \pm 0.15) \cdot 10^{-3} eV^2$& $(2.1 -
2.8) 10^{-3} eV^2$\\
 $tan^2 \theta_{12} = 0.484 \pm 0.048$& $31^0 < \theta_{12} < 39^0$\\
 $sin^2 2\theta_{23} = 1.02 \pm 0.04$& $ 37^0 < \theta_{23} < 53^0$\\
$^*\quad$ $\sin^2 2\theta_{13} = 0.11$ ($ \theta_{13}=10^0$)&\\
\hline
\end{tabular}\label{Tab}
\end{center}

{\small Table 1. Experimental constraints on the parameters of
active neutrinos \cite{Strumia}, $^*$ --- results of T2K
Collaboration \cite{T2Ktheta13}:
 $0.03<\sin^2 2\theta_{13}<0.28$ in the case of the normal hierarchy  and $0.04<\sin^2 2\theta_{13}<0.34$ in the case of the inverted hierarchy.}

\phantom{sss}

On the other hand, from the "see-saw"\, formula (in the
approximation when the elements of the first column of the Yukawa
matrix are neglected and  $M\gg m_{ij}$) one can immediately obtain,
that the mass of the lightest sterile neutrino is zero and the mass
matrix of the active neutrinos has the form \cite{sb}
\begin{equation}
\label{system} [M_{light}]_{\alpha\beta}=-\frac{v^2}{2M}(h_{\alpha
2}h_{\beta 3}+h_{\alpha 3}h_{\beta 2}),
\end{equation}
and its eigenvalues is
\begin{equation}\label{m2m3}
m_a=0,\quad m_{\tiny \left(\!\!\!\begin{array}{c}
b\\c\end{array}\!\!\!\right)}=\frac{v^2(F_2F_3\mp |h^+
h|_{23})}{2M},
\end{equation}
where $F_I^2=(h^+ h)_{II}$, $m_a$ is the mass of the lightest active
neutrino, $m_c$ is the mass of the heaviest active neutrino. The sum
over the neutrino masses is given by
\begin{equation}
\label{sum} \frac{v^2F_2F_3}{M}=\sum_{i=1}^3m_{i}.
\end{equation}

The system \eqref{system} has infinite number of solutions. Indeed,
the replacement  $h_{\alpha2}\rightarrow zh_{\alpha2}$,
$h_{\alpha3}\rightarrow h_{\alpha3}/{z}$ ($z$ is an arbitrary
complex number) does not change the system. Then one can define the
real quantity $\varepsilon$
\begin{equation}\label{epsilondef}
\varepsilon={F_3}/{F_2},\quad\varepsilon=|z|.
\end{equation}
as an independent parameter of the model.

As it was shown in \cite{GorkVil}, the system \eqref{system}  has
good solutions for  ratios of the elements of second column of the
Yukawa matrix:
\begin{equation}\label{solution}
\left\{\begin{array}{c}
\vspace{0.5em}A_{12}=\displaystyle{\frac{M_{12}}{M_{22}}\left(1\pm\sqrt{   1-\frac{M_{11}M_{22}}{M_{12}{}^{2}} }  \right)}\\
\vspace{0.5em}A_{13}=\displaystyle{\frac{M_{13}}{M_{33}}\left(1\pm\sqrt{1-\frac{M_{11}M_{33}}{M_{13}{}^{2}} }  \right)}\\
A_{23}=\displaystyle{\frac{M_{23}}{M_{33}}\left(1\pm\sqrt{1-\frac{M_{22}M_{33}}{M_{23}{}^{2}}}\right)}
\end{array}\right.
\end{equation}
where $A_{12}={h_{12}}/{h_{22}}$, $A_{13}={h_{12}}/{h_{32}}$,
$A_{23}={h_{22}}/{h_{32}}$ and $M_{IJ}$ is elements of $M_{light}$
matrix. The ratios of the third column elements of the Yukawa matrix
are expressed through the $A_{ij}$ elements:
\begin{equation}\label{dop}
\frac{h_{23}}{h_{13}}=A_{12}\frac{M_{22}}{M_{11}};\quad
\frac{h_{33}}{h_{13}}=A_{13}\frac{M_{33}}{M_{11}}.
\end{equation}

Though formally there are eight different choices for the solutions
\eqref{solution}, only four are independent. For example, if we fix
the sign before the square roots in the expressions for $A_{12}$ and
$A_{13}$ then $A_{23}$ is unambiguously determined by the relation
\begin{equation}\label{a10}
A_{23}=A_{13}/A_{12}.
\end{equation}

The solutions \eqref{solution} allow one to find the ratios of the
elements of Yukawa matrix \cite{GorkVil}:
\begin{equation}
\label{haI1}
\frac{\left(h_{12};h_{22};h_{32}\right)}{F_2}=\frac{e^{iarg(h_{12})}}{\sqrt{1+|A_{12}|^{-2}+|A_{13}|^{-2}}}\left(1;A_{12}^{-1};A_{13}^{-1}\right)
\end{equation}
\begin{equation}
\label{haI2}
\frac{\left(h_{13};h_{23};h_{33}\right)}{F_3}=\frac{e^{iarg(h_{13})}}{\sqrt{1+|A_{12}\frac{M_{22}}{M_{11}}|^{2}+|A_{13}\frac{M_{33}}{M_{11}}|^{2}}}\left(1;A_{12}\frac{M_{22}}{M_{11}};A_{13}\frac{M_{33}}{M_{11}}\right),
\end{equation}
where phases of $h_{12}$, $  h_{13}$ are connected by condition
\begin{equation}
arg(h_{12})+arg(  h_{13})=arg(M_{11}).
\end{equation}
This is the exact solution of \eqref{system} that  definitely
expresses ratio of the elements of the Yukawa matrix via parameters
of the active neutrino mass matrix. For fixed values of the active
neutrino parameters there are only two choices for placing of the
signs in the expressions
 for $A_{12},A_{13},A_{23}$  \eqref{solution} which are not inconsistent
with condition \eqref{a10}. These two variants are distinguished
from each other by simultaneous replacement of the sign in front of
square roots in the expressions for $A_{12},A_{13}, A_{23}$. It  can
be shown that such replacement of the signs leads to interchanging
and conjugating of the ratios of elements of the second and the
third columns of the Yukawa matrix, notably
$h_{22}/h_{12}\leftrightarrow h^*_{23}/h^*_{13}$,
$h_{32}/h_{12}\leftrightarrow h^*_{33}/h^*_{13}$  \cite{GorkVil}.

As it was announced in Introduction, only the two heavier sterile
neutrino take part in the production of the leptonic asymmetry.
Therefore we will exclude the lightest sterile neutrino from
consideration, so hereinafter indexes $I,J$ take the value 2 or 3
referring to the two heavy sterile neutrinos. In this case there are
11 additional parameters in the $\nu MSM$ as compared with SM. Seven
of them we will identify with the elements of the active neutrino
mass matrix ($m_2$, $m_3$, $\theta_{12}$, $\theta_{13}$,
$\theta_{23}$, $\delta$, $\alpha_2$). The other 4 we will define as
follows: the average mass of two heavier sterile neutrinos
$M=\frac{M_2+M_3}{2}$, their mass splitting $\Delta
M=\frac{M_3-M_2}{2}$, the parameter $\varepsilon$ and the phase
$\xi=arg(h_{12})$.

Thus, we can parameterize the Lagrangian \eqref{nongiag1} in the
following way:
\begin{multline}
\label{lagparam} \mathcal L=\left(\frac{M\sum
m_{\nu_i}}{v^2}\right)^{\frac{1}{2}}\left[\frac{1}{F_2\sqrt{\varepsilon
}}h_{\alpha 2}\bar L_{\alpha}\tilde N_\alpha+\frac{\sqrt{\varepsilon
}}{F_3}h_{\alpha 3}\bar L_{\alpha}\tilde
N_3\right]\tilde\Phi-\\-M\bar{\tilde N}_2^c\tilde
N_3-\frac{1}{2}\Delta M\left(\bar{\tilde{ N_2^c}}\tilde
N_2+\bar{\tilde{ N_3^c}}\tilde N_3\right)+h.c.,
\end{multline}
where  $a_{\alpha I}={h_{\alpha I}}/{F_I}$  are defined by equations
\eqref{haI1} and \eqref{haI2}.

Lagrangian \eqref{lagdiag0} can be written in another basis, namely
when the mass matrix of sterile right-handed neutrinos is diagonal.
In this case the Lagrangian is
\begin{equation}\label{diag}
\mathcal{L}^{ad}=-g_{\alpha I}\bar{L}_\alpha {N'_I}\tilde
{\Phi}-\frac{M_I}{2}\bar{N}'{}_I^cN'_I+h.c.,
\end{equation}
where ${N'_I}$ are  right-handed neutrinos and $g_{\alpha I}$ are
elements of the Yukawa matrix in this basis.

Transition from presentation of Lagrangian \eqref{lagdiag0} in gauge
and mass basis can be made with unitary transformation that
transfers mass matrix of right-handed neutrino  to diagonal form
\cite{Tibor,sb}:
\begin{equation}\label{transform}
V^*\left(\begin{array}{cc}\Delta M&M\\M&\Delta
M\end{array}\right)V=\left(\begin{array}{cc}M-\Delta M&0\\0&M+\Delta
M\end{array}\right);\quad
V=\frac1{\sqrt{2}}\left(\begin{array}{cc}-i&i\\1&1\end{array}\right).
\end{equation}
So, the transition can be made by
\begin{equation}\label{transition}
\tilde N_I=V_{IJ} N'{}_J,\quad g_{\alpha I}=h_{\alpha J} V_{JI}.
\end{equation}
With help of this relations it will be useful to express Lagrangian
\eqref{lagparam} in terms of right-handed neutrino functions of
Lagrangian \eqref{diag}
\begin{multline}\label{lagdiag1}
\mathcal{L}^{ad}=-\left(\frac{M\sum
m_{\nu_i}}{2v^2}\right)^{\frac{1}{2}}\left[\left(\frac{ia_{\alpha
2}}{\sqrt{\varepsilon }}-i\sqrt{\varepsilon }a_{\alpha 3}\right)\bar
L_{\alpha}  N'{}_2+\left(\frac{a_{\alpha 2}}{\sqrt{\varepsilon
}}+\sqrt{\varepsilon }a_{\alpha 3}\right)\bar L_{\alpha}
N'{}_3\right]\tilde\Phi-\\-\frac{1}{2}\left(M-\Delta M\right)\bar{
{N'{}_2^c}}  N'{}_2-\frac{1}{2}(M+\Delta M)\bar{ {N'{}_3^c}}
N'{}_3.
\end{multline}
After comparing \eqref{lagdiag1} and \eqref{diag} one can express
Yukawa couplings in different presentations
\begin{align}
&g_{\alpha 2}=\left(\frac{M\sum_i
m_{\nu_i}}{2v^2}\right)^{\frac{1}{2}}\left(\frac{ia_{\alpha
2}}{\sqrt{\varepsilon }}-i\sqrt{\varepsilon }a_{\alpha 3}\right),\\
&g_{\alpha 3}=\left(\frac{M\sum_i
m_{\nu_i}}{2v^2}\right)^{\frac{1}{2}}\left(\frac{a_{\alpha
2}}{\sqrt{\varepsilon }}+\sqrt{\varepsilon }a_{\alpha 3}\right).
\end{align}

The mass eigenstates  neutrinos for Lagrangian with the mass matrix
$M^{DM}$ \eqref{dm2} can be easily expressed through the states of
neutrino of Lagrangian \eqref{diag}, particularly:
\begin{equation}\label{through}
N^c=\left(1-\frac{1}{2}\varepsilon\varepsilon^+\right)N'{}^c+\varepsilon\nu_{L}\simeq
N'{}^c+\varepsilon\nu_{L},
\end{equation}
where $N$ are mass eigenstates of the right-handed neutrinos in
which they are produced and decay, $\nu_{L}$ are the active
neutrinos of the SM in flavor basis,
\begin{equation}\label{theta}
\varepsilon_{\alpha I}\equiv\Theta_{\alpha
I}=\frac{v}{\sqrt{2}}\frac{g_{\alpha I}}{M_I}
\end{equation}
is the mixing angle ($\varepsilon_{\alpha I} \ll 1$).

\section{The computation of the leptonic asymmetry.}$\qquad$
$\qquad$ \label{rozdil3}

As it was pointed in Section \ref{intro}, leptonic asymmetry in the
$\nu MSM$ is generated due to decays of the heavier sterile
neutrinos on SM particles. At temperature $T\ll T_{EW}$ the
interaction of the sterile neutrinos with SM particles via neutral
Higgs field can be neglected. The only possible way of interaction
of the sterile neutrino with matter is through the mixing with
active neutrinos \eqref{through}.

For the sterile neutrino with the mass $m_\pi<M_I<$ 2 GeV  the
channels for the decay into two-body final state are:
\begin{equation}\label{decay}
N_I\rightarrow \pi^0\nu_\alpha,\pi^+e_\alpha^-,\pi^-e_\alpha^+,
K^+e_\alpha^-, K^-e_\alpha^+,\eta\nu_\alpha,\eta
'\nu_\alpha,\rho^0\nu_\alpha,\rho^+e_\alpha^-,\rho^-e_\alpha^+.
\end{equation}
The channel of decay $N_{2,{3}}\rightarrow N_1+...$ is strongly
suppressed because of the small Yukawa coupling constants of $N_1$.
The decay of the sterile neutrino into the $K^0$ state is forbidden,
because the composition of $K^0$ ($d\bar s$) can not be obtained by
decay of $Z$-boson.

The three-body final state can be safely neglected and also the many
hadron final state \cite{Gorbunov}. This last decay channels
contribute for less than 10\% for $M_I < 2$ GeV. For $m_\pi < M_I <
2$ GeV the decays into D-meson can also be neglected because its
mass is not much smaller than 2 GeV.

Let us consider the decay of the sterile neutrino in the $\nu MSM$.
Sterile neutrino oscillates into active neutrino that decay into
$Z$-bozon and active neutrino (or $W^\pm$-boson and charged lepton)
in accordance with the SM. $Z$-boson (or $W^\pm$-boson) hereafter
decays into quark-antiquark pair, see Fig.1. Since kinetic energy of
this quarks are small enough the quark pair will form a bound state.
Since $M_I<2$ GeV $\ll M_{Z(W)}$ we can use low energy Fermi theory
and shrink the heavy boson propagator into an effective vertex and
use for final state a meson, see  Fig.2.

\begin{figure}[h]
\begin{tabular}{c}
\includegraphics[width=150mm]{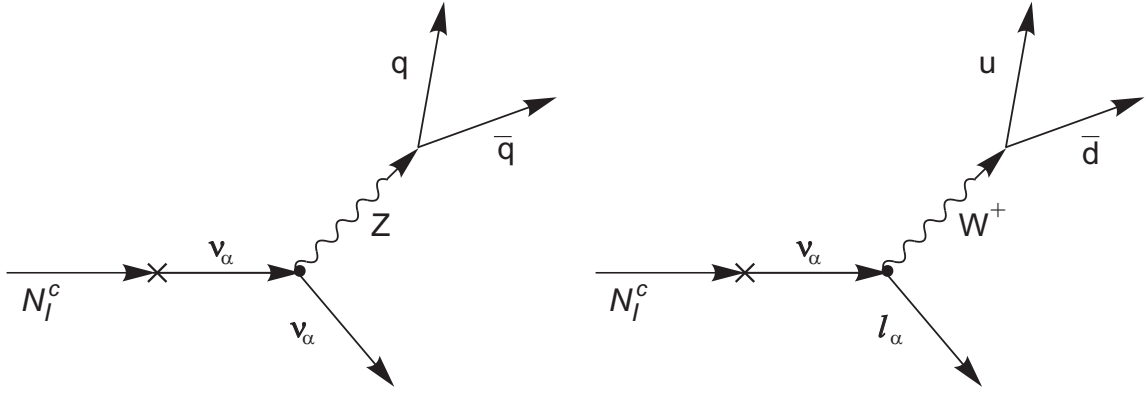}
\end{tabular}
\caption{The decay of a sterile neutrino via $Z$-boson and
$W^+$-boson (the cross on line of a sterile neutrino means an
oscillation of a sterile to an active neutrino).}
\end{figure}

\begin{figure}[b]
\begin{tabular}{c}
\qquad\qquad\qquad\qquad\includegraphics{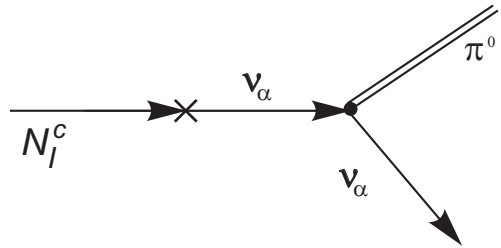}
\end{tabular}
\caption{Effective low-energy decay of a sterile neutrino into
$\pi^0$ meson and active neutrino.}
\end{figure}

The process of sterile neutrino decay into charged lepton and
charged meson through $W^\pm$-boson  is described by charged current
interaction
\begin{equation}\label{epi1}
\mathcal{L}_C=\frac{G_F}{\sqrt{2}}\left(j_\nu^{CC}\right)^+j^{\nu\,CC},
\end{equation}
where $j_\nu^{CC}=j_\nu^{l\,CC}+j_\nu^{h\,CC}$ is charged lepton and
hadron current,
\begin{equation}\label{epi2}
j_\nu^{l\,CC}=\sum_\alpha {\bar
e}_\alpha\gamma_\nu(1-\gamma^5)\nu_\alpha, \quad
j_\nu^{h\,CC}=\sum_{n,m} V_{n,m}^*{\bar
d}_m\gamma_\nu(1-\gamma^5)u_n.
\end{equation}
The indices $m,n$ run over the quark generation, $\alpha=e,\mu,\tau$
and $V$ is Kabbibo-Kobayashi-Maskawa (CKM) matrix. Similarly, the
process of sterile neutrino decay into active neutrino  and neutral
meson through $Z$-boson  is described by neutral current interaction
\begin{equation}\label{epi4}
\mathcal{L}_N=\sqrt{2}{G_F}\left(j_\nu^{NC}\right)^+j^{\nu\,NC},
\end{equation}
where $j_\nu^{NC}=j_\nu^{l\,NC}+j_\nu^{h\,NC}$ is active neutrino
and hadron current,
\begin{equation}\label{epi5}
j_\nu^{l\,NC}=\sum_\alpha {\bar
\nu}_\alpha\gamma_\nu\frac{1-\gamma^5}2\nu_\alpha, \quad
j_\nu^{h\,NC}=\sum_{f} {\bar
f}\gamma_\nu\left(t_3^f(1-\gamma^5)-2q_f\sin^2\theta_W\right)f,
\end{equation}
where sum over $f$ means sum over all quarks, $t^f_3$ -- is the weak
isospin of the quark, $q_f$ --- is the electric charge of quark in
proton charge units, notably $t^f_3=1/2, q_f=+2/3$ for $u,c,t$ and
$t^f_3=-1/2, q_f=-1/3$ for $d,s,b$ quarks.

The matrix element corresponding to Feynman diagram of sterile
neutrino decay (see, e.g., Fig.1,2) can be obtained from the
interactive effective Lagrangian \cite{Tibor}. For example,
effective Lagrangian of decay of $I$ sterile neutrino into the
$\pi^\pm,\pi^0$ final states  is:
\begin{multline}\label{Lpi}
 \mathcal
L_{eff}^{\pi}=\frac{G_F}{2}M_If_{\pi}\Theta_{\alpha I}\bar
\nu_{\alpha}(1+\gamma_5)N_I \pi^0+\\
+\left[\frac{G_F}{\sqrt{2}}M_If_{\pi}V_{ud}\Theta_{\alpha I}\bar
e_{\alpha}\left((1+\gamma_5)-\frac{m_{\alpha}}{M_I}(1-\gamma_5)\right)N_I \pi^-+h.c.\right],
\end{multline}
where $G_F$ is Fermi coupling constant, $M_I$ is the mass of
$I$-sterile neutrino, $m_\alpha$ is the mass of the charged lepton
of $\alpha$ generation,  $f_\pi$ is the $\pi$-meson decay constant
that is defined as
\begin{equation}\label{constf}
 \langle\pi^+|\bar u(1+\gamma^5)\gamma_\nu d|0\rangle=-f_\pi\cdot
(p_\pi)_\mu,
\end{equation}
where $p_\pi$ is the pion 4-momentum.

 The leptonic asymmetry $\epsilon$ can be defined as
\begin{equation}\label{epsilon}
\epsilon=\frac{\Gamma_{N\rightarrow l}-\Gamma_{N\rightarrow \bar
l}}{\Gamma_{N\rightarrow l}+\Gamma_{N\rightarrow \bar l}}\,,
\end{equation}
where $\Gamma_{N\rightarrow l}$ is the total decay rate of sterile
neutrinos into leptons and $\Gamma_{N\rightarrow \bar l}$ is the
total decay rate of sterile neutrinos into antileptons.

At tree level the decay rates of the sterile neutrinos into leptons
and antileptons are equal. Therefore we must compute the one loop
diagrams, see Fig.3. In the case of nearly degenerated sterile
neutrinos the contribution from the diagrams presented at
Fig.3\textit{b}) can be neglected as compared with
 diagrams  presented at Fig.3\textit{a}). Indeed the
propagator of the sterile neutrino in the diagrams \textit{a)} type
is proportional to $1/\Delta M$ in the center of mass frame. The
leading order contribution to the leptonic asymmetry comes from
interference between one-loop diagrams and tree-level diagrams
\cite{Davidson}. In this case $\Gamma_{N\rightarrow
l}-\Gamma_{N\rightarrow \bar l}\sim \Theta^4$ and
$\Gamma_{N\rightarrow l}+\Gamma_{N\rightarrow \bar l}\sim \Theta^2$,
the leptonic asymmetry  is suppressed.

\begin{figure}[t]
\begin{tabular}{c}
\qquad\qquad\qquad\qquad\includegraphics[width=90mm]{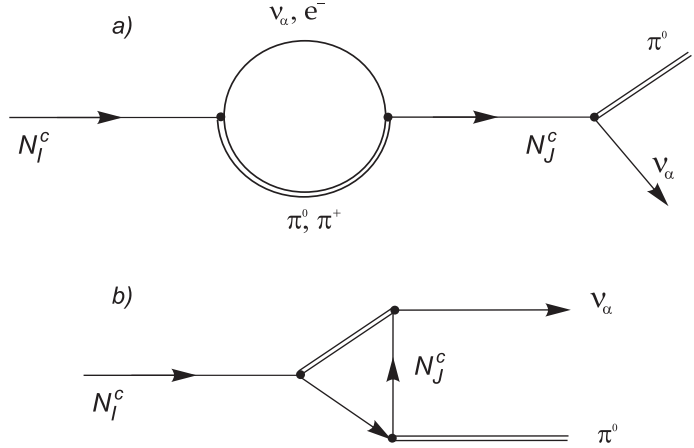}
\end{tabular}
\caption{Example of one-loop diagrams of the decay $N_I\rightarrow
\nu_\alpha\pi^0$.}
\end{figure}

In our case, when the mass splitting between the two heavier sterile
neutrinos is very small and it is of the same order as their decay
rate (we obviously will see it later), the oscillations between
$N_I$ and $N_J$ are important, see Fig.4. So, the corresponding mass
eigenfunctions are no longer the $N_I$ states, but a mixture of
them, namely $\psi_I$ \cite{Flanz,Tibor}. It is these physical
eigenstates which evolve in time with a definite frequency. The
subsequent decay of these fields will produce the desired lepton
asymmetry
\begin{equation}\label{Delta}
\Delta=\frac{\Gamma_{\psi\rightarrow l}-\Gamma_{\psi\rightarrow \bar
l}}{\Gamma_{\psi\rightarrow l}+\Gamma_{\psi\rightarrow \bar l}}\,,
\end{equation}
where $\Gamma_{\psi\rightarrow l}$ and  $\Gamma_{\psi\rightarrow
\bar l}$ are the total decay rates of the sterile neutrino mass
eigenfunctions $\psi_I$ into leptons and  antileptons
correspondingly. In this case the leading order contribution to the
leptonic asymmetry comes from tree-level diagrams.

\begin{figure}[t]
\begin{tabular}{c}
\qquad\qquad\qquad\qquad\includegraphics[width=90mm]{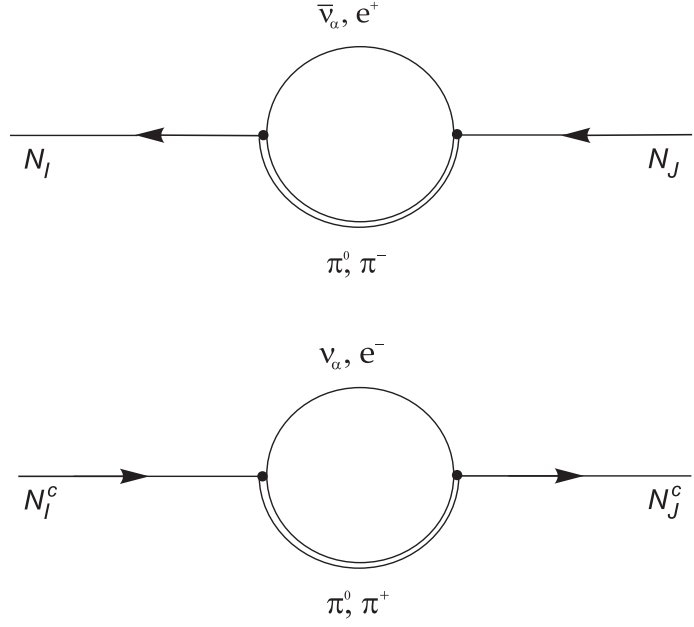}
\end{tabular}
\caption{Contributions to the effective Hamiltonian.}
\end{figure}

In general case the correct description of the processes can be made
in  frame of the density matrix formalism, see, e.g., \cite{Shap1}.
We will follow a simpler way by considering a non-hermitian
Hamiltonian. The effective Hamiltonian in the basis of $\{N_2,N_3\}$
is the $H=H_0+\Delta H$, where $H_0$ is the diagonal Hamiltonian of
equal mass particle
\begin{equation}\label{H0}
H_0=\left(\begin{array}{cc}M&0\\0&M\end{array}\right).
\end{equation}
The corrections to this Hamiltonian are given by the one-loop
diagrams, see Fig.4:
\begin{equation}\label{DeltaH}
\Delta H=\left(\begin{array}{cc}-\Delta
M-\frac{i}{2}\Gamma_2&-\frac{i}{2}\Gamma_{23}\\-\frac{i}{2}\Gamma_{23}&\Delta
M-\frac{i}{2}\Gamma_3\end{array}\right).
\end{equation}
The dispersive part of these diagrams can be absorbed in the mass
renormalization of the fields \cite{Flanz} and it brings to
appearance of the mass splitting $\Delta M$. The absorptive part of
the diagrams 
will define total decay
rates of the sterile neutrino $\Gamma_I$ and the rate of oscillation
between sterile neutrinos $\Gamma_{23}$.

Total  rates of $I$-sterile neutrino decays into charged mesons and
leptons of $\alpha$-generation are
\begin{multline}\label{GP+-}
\Gamma_I^{\,\alpha\pi^\pm}=\Gamma(N_I\rightarrow
\pi^\pm+l_\alpha^{\mp})=\\=\frac{G_F^2f_\pi^2|V_{ud}|^2M^3}{8\pi}|\Theta_{\alpha
I}|^2S(M,m_\alpha,m_\pi)\left[\left(1-\frac{m_\alpha^2}{M^2}\right)^2-\frac{m_\pi^2}{M^2}\left(1+\frac{m_\alpha^2}{M^2}\right)\right],
\end{multline}
\begin{multline}\label{GK}
\Gamma_I^{\,\alpha K}=\Gamma(N_I\rightarrow
K^\pm+l_\alpha^{\mp})=\\=\frac{G_F^2f_K^2|V_{us}|^2M^3}{8\pi}|\Theta_{\alpha
I}|^2S(M,m_\alpha,m_K)\left[\left(1-\frac{m_\alpha^2}{M^2}\right)^2-\frac{m_K^2}{M^2}\left(1+\frac{m_\alpha^2}{M^2}\right)\right],
\end{multline}
\begin{multline}
\Gamma_I^{\,\alpha\rho^\pm}=\Gamma(N_I\rightarrow
\rho^\pm+l^\mp_{\alpha})=\\=\frac{G_F^2g_\rho^2|V_{ud}|^2M^3}{4\pi
m_\rho^2}|\Theta_{\alpha
I}|^2S(M,m_\alpha,m_\rho)\left[\left(1-\frac{m_\alpha^2}{M^2}\right)^2+\frac{m_\rho^2}{M^2}\left(1+\frac{m_\alpha^2-2m_\rho^2}{M^2}\right)\right],
\end{multline}
where
\begin{equation}\label{S}
   S(M_I,m_\alpha,m)=\sqrt{\left(1-\frac{(m-m_\alpha)^2}{M_I^2}\right)\left(1-\frac{(m+m_\alpha)^2}{M_I^2}\right)},
\end{equation}
and values of decay constants and elements of CKM matrix are given
in \cite{PG}: $f_\pi=0.131$ GeV, $f_K=0.16$ GeV, $g_\rho=0.102$
GeV$^2$, $|V_{ud}|=0.97$, $|V_{us}|=0.23$.

Total  rates of $I$-sterile neutrino decays into neutral mesons and
active neutrinos are
\begin{equation}\label{GP0}
\Gamma_I^{\,\alpha\pi^0}=\Gamma(N_I\rightarrow \pi^0+\nu_\alpha)=
\frac{G_F^2f_\pi^2M^3}{16\pi}|\Theta_{\alpha
I}|^2\left(1-\frac{m_\pi^2}{M^2}\right)^2,
\end{equation}
\begin{equation}
\Gamma^{\,\alpha\rho^0}_I=\Gamma(N_I\rightarrow
\rho_0+\nu_{\alpha})=\frac{G_F^2g_\rho^2M^3}{8\pi
m_\rho^2}|\Theta_{\alpha
I}|^2\left(1+2\frac{m_\rho^2}{M^2}\right)\left(1-\frac{m_\rho^2}{M^2}\right)^2,
\end{equation}
\begin{equation}
\Gamma_I^{\,\alpha\eta}=\Gamma(N_I\rightarrow
\eta+\nu_{\alpha})=\frac{G_F^2f_{\eta}^2M^3}{16\pi}|\Theta_{\alpha
I}|^2\left(1-\frac{m_{\eta}^2}{M^2}\right)^2,
\end{equation}
\begin{equation}\label{eta'}
\Gamma_I^{\,\alpha\eta'}=\Gamma(N_I\rightarrow
\eta'+\nu_{\alpha})=\frac{G_F^2f_{\eta'}^2M^3}{16\pi}|\Theta_{\alpha
I}|^2\left(1-\frac{m_{\eta'}^2}{M^2}\right)^2,
\end{equation}
where $f_\eta=0.156$ GeV, $f_\eta'=-0.058$ GeV \cite{PG}.

As one can see the decay rates into $\rho^\pm,\rho^0$ mesons are
slightly different because they are  vector mesons. The adduced
decay rates \eqref{GP0} -- \eqref{eta'}  were obtained in
\cite{Johnson,Gorbunov}. The total decay rate of sterile neutrino
decay into mesons and  leptons  is sum of the rates over all decay
channels $\Lambda$ \eqref{decay} and over  leptonic generation:
\begin{equation}\label{totalrate}
\Gamma_{I}=\sum_{\alpha,\Lambda}\Gamma_I^{\,\alpha\Lambda}\,\Theta(y_{\alpha\Lambda}),
\end{equation}
where $y_{\Lambda}$ is the difference of the $I$ sterile neutrino
mass and total mass of all final particles of the decay channel
$\Lambda$; $\Theta(x)$ is the usual Heaviside function.
The rate of oscillation between $I$ and $J$ sterile neutrinos
($\Gamma_{IJ}$) can be expressed through the decay rates
\begin{equation}\label{IJ}
 \Gamma_{IJ}=\sum_{\alpha,\Lambda}
\frac{{\rm Re}(\Theta_{\alpha I}\Theta_{\alpha
J}^*)}{|\Theta_{\alpha
I}|^2}\,\Gamma_I^{\,\alpha\Lambda}\,\Theta(y_{\alpha\Lambda}).
\end{equation}

The eigenvalues and corresponding eigenfunctions of the
non-hermitian Hamiltonian $H=H_0+\Delta H$ are given by
\begin{align}\label{evalues}
&
\omega_2=M-\frac{i}{4}(\Gamma_2+\Gamma_3)-\frac{1}{4}c,\qquad\psi_2=\frac{1}{\sqrt{N}}\left(\begin{array}{c}B\\1\end{array}\right),\\
&\omega_3=M-\frac{i}{4}(\Gamma_2+\Gamma_3)+\frac{1}{4}c,\qquad
\psi_3=\frac{1}{\sqrt{N}}\left(\begin{array}{c}1\\-B\end{array}\right),
\end{align}
where $N$ is a normalization factor and $$c=\sqrt{(4\Delta
M-i(\Gamma_3-\Gamma_2)^2-4(\Gamma_{23})^2},\quad B=(4i\Delta
M+(\Gamma_3-\Gamma_2)+ic)/({2\Gamma_{23}}).$$

It should be noted the sterile neutrinos are not initially in the
state $\psi_2$ and $\psi_3$, but in the state $N_2$ and $N_3$.  The
fact is that sterile neutrino where in thermal equilibrium  before
they propagated freely. The equilibrium was maintained by the weak
interaction between the sterile neutrinos and particles in the
background. The weak interaction eigenstates are $N_2$ and $N_3$,
therefore at the beginning the sterile neutrinos are in the state
$N_2$ or $N_3$. In general the initial state of sterile neutrino is
the superposition of $N_2$ and $N_3$ states and can be described by
a density matrix:
\begin{equation}
\label{roini}
\hat\rho_{initial}=\hat\rho(t=0)=\sum_{I=2,3}\alpha_I|N_I(0)\rangle\langle
N_I(0)|,
\end{equation}
where $\alpha_2+\alpha_3=1$. It was shown in \cite{Tibor}  that
leptonic asymmetry dependence on parameter $\alpha_I$ can be
neglected. We confirmed this statement and, hereafter, we will
consider the symmetric initial state $\alpha_2=\alpha_3={1}/{2}$.

The time evolution of the density matrix can be obtain in a simple
way. Since
\begin{equation}
\label{uperetv} |\psi_I\rangle=U_{IJ}|N_J\rangle,
\end{equation}
where
\begin{equation}
\label{u}
U=\frac1{\sqrt{N}}\left(\begin{array}{cc}B&1\\1&-B\end{array}\right),
\end{equation}
the time evolution of $|N_I\rangle$ state is known
\begin{equation}
\label{evolstates}
|N_I(t)\rangle=U^{-1}_{IK}e^{-i\omega_Kt}|\psi_K(0)\rangle=U^{-1}_{IK}e^{-i\omega_Kt}U_{KJ}|N_J(0)\rangle=R_{IJ}|N_J(0)\rangle.
\end{equation}
Thus
\begin{equation}\label{roT}
\hat\rho(t)=\frac{1}{2}\sum_{I,J,K=2}^3R_{IK}(t)^\ast
R_{IJ}(t)|N_J(0)\rangle\langle
N_K(0)|=\frac{1}{2}\sum_{J,K=2}^3(R^\dag
R)_{KJ}|N_J(0)\rangle\langle N_K(0)|.
\end{equation}

The average production rate of leptons is  given by
\begin{multline}
\Gamma=\!\!\int_0^\infty \!\!\!\!\!dt\!\!\int
\!\!\!d\Pi_2\sum_lTr\left[|l\rangle\langle
l|\hat\rho(t)\right]=\frac{1}{2}\int_0^\infty \!\!\!\!\!dt\!\!\int
\!\!\!d\Pi_2\, Tr\left[\sum_{l,K,J}(R^\dag R)_{KJ}\langle
l|N_J(0)\rangle\langle
N_K(0)|l\rangle\right]\!=\\=\frac{1}{2}\int_0^\infty
\!\!\!\!\!dt\!\!\int \!\!\!d\Pi_2\sum_{l,J,K}(R^\dag
R)_{KJ}A_{Jl}A_{Kl}^\ast,
\end{multline}
where sum over $l$ means sum over all leptons generations and
include charged leptons and active neutrinos, $\langle
l|N_J(0)\rangle=A_{Jl}$ is the transition amplitude of the decay of
$I$ sterile neutrino
 into a lepton at tree level that includes all possible channels of reaction, and $d \Pi_2$ is the differential 2-body
phase space
$$d\Pi_2=\frac{d^3q}{(2\pi)^32E_q}\frac{d^3k}{(2\pi)^32E_k}(2\pi)^4\delta^4(p-q-k),$$
where $p,q,k$ are 4-momentums of initial and final particles in
decay.

Similarly the production rate of antileptons is
\begin{equation}
\bar \Gamma=\frac{1}{2}\int_0^\infty \!\!\!\!\!dt\!\!\int
\!\!\!d\Pi_2\sum_{l,J,K}(R^\dag R)_{KJ}A_{Jl}^\ast A_{KJ}.
\end{equation}

The measure of the leptonic asymmetry is given by
\begin{multline}
\!\Delta\!=\!\frac{\Gamma\!-\!\bar \Gamma}{\Gamma\!+\!\bar
\Gamma}\!=\! \frac{\int dt\int d\Pi_2Im((R^\dag
R)_{32})Im(A_{2l}^\ast A_{3l})}{\int dt\int d\Pi_2((R^\dag
R)_{22}|A_2l|^2\!+\!(R^\dag R)_{33}|A_3l|^2\!+\!2Re(A_{2l}^\ast
A_{3l})Re(R^\dag R)_{23})}
\end{multline}
The integration over $d\Pi_2$ gives \cite{Tibor}:
\begin{equation}\label{assym}
\Delta=\frac{\int dt Im((R^\dag
R)_{32})\sum_{\alpha}Im(\Theta_{\alpha 2}^\ast \Theta_{\alpha
3})V_{\alpha}}{\int dt\sum_{\alpha}((R^\dag R)_{22}|\Theta_{\alpha
2}|^2+(R^\dag R)_{33}|\Theta_{\alpha 3}|^2+2Re(\Theta_{\alpha
2}^\ast \Theta_{\alpha 3})Re(R^\dag R)_{23})V_{\alpha}},
\end{equation}
where $V_\alpha$ is defined via sum over all possible channels of
sterile neutrino decays into  leptons of  generation $\alpha$
\begin{equation}\label{v}
V_\alpha=\sum_{\Lambda}
\frac{\Gamma_I^{\,\alpha\Lambda}}{|\Theta_{\alpha
I}|^2}\,\Theta(y_{\alpha\Lambda}).
\end{equation}

\section{The restrictions on the parameters of the $\nu MSM$}\qquad
\label{rozdil4}

As it was pointed in Section \ref{intro}, the leptonic asymmetry of
the Universe has to be constrained by condition \eqref{asymmetry} at
the moment of the beginning  of the DM particles production. It
allows us to constrain parameters of the $\nu MSM$. To do it, we can
construct the leptonic asymmetry \eqref{assym} as function of only
three parameters of $\nu MSM$: $M$,  $\Delta M$, $\varepsilon$.

We do it in the following way. Leptonic asymmetry  function
\eqref{assym} is maximized over phases $\delta$, $\alpha_2$, $\xi$
(and $\alpha_1$ in case of the inverted hierarchy) and is taken at
central value of active neutrino mass matrix parameters\footnote{In
case of the normal hierarchy we have $m_1=0$, $m_2=\sqrt{\Delta
m_{21}^2}=0.009\,eV$, $m_3=\sqrt{|\Delta m^2_{23}|+\Delta
m_{21}^2}=0.05\,eV$. In case of inverted hierarchy we have
$m_1=\sqrt{|\Delta m^2_{23}|-\Delta m_{21}^2}=0.048\,eV$,
$m_2=\sqrt{|\Delta m^2_{23}|}=0.049\,eV$, $m_3=0$.}, see Tab.1. This
function contains dependence on ratios of the Yukawa matrix elements
in mixing angle $\Theta_{\alpha I}$ \eqref{theta} that can be
expressed through solutions \eqref{solution} with two possible
choice of sign consistent with condition \eqref{a10}. So far as the
relation for leptonic asymmetry \eqref{assym}  has no symmetry for
interchanging and conjugating of the ratios of elements of the
second and the third columns of the Yukawa matrix we have to
consider two variants of the solutions.
 For fixed values of the
mixing angles and phases we will designate allowed solution of
\eqref{solution}
 with 2 or more sign $(+)$ as solution of A
type, and, vice-versa, the solution with 2 or more sign $(-)$ we
will designate as solution of B type. It should be noted that our
results \eqref{haI1}, \eqref{haI2} for B type of solution coincide
with results of \cite{Sh3} where the ratios of the elements were
obtained in the particular case $\theta_{13}\rightarrow 0$,
$\theta_{23}\rightarrow\pi/4$. We separately consider the case of
$\theta_{13}=0$ and $\theta_{13}=10^{\rm o}$ also.

Thereby we construct allowed regions ($\Delta>10^{-3}$) in plane of
parameters $\Delta M$ and $\varepsilon$ at fixed values of $M$.

For the case of the normal hierarchy the deference between the case
of $\theta_{13}=0$ or $\theta_{13}=10^{\rm o}$ and the case of
solution of A or B types is not essential, so we illustrate allowed
regions with help of only one figure on Fig.5. For the case of the
inverted hierarchy, the difference between the case of solution of A
or B types is not essential, but the cases of $\theta_{13}=0$ and
$\theta_{13}=10^{\rm o}$ are substantially different. So we
illustrate allowed regions with help of  two figures on Fig.6.

It should be noted that we investigated form of the allowed regions
not only for the central value of  $\theta_{13}$ angle, but for
range given by data of \cite{T2Ktheta13}. We conclude that in case
of the normal hierarchy the regions are almost not sensitive to
value of $\theta_{13}$ in range $0<\theta_{13}<16^{\rm o}$. In case
of the inverted hierarchy it is true for the regions on Fig6
\textit{b}) and $\theta_{13}<18^{\rm o}$, but for $\theta_{13}=0$
the allowed regions are appreciably different.

Also we illustrate  regions where maximum of $\Delta$ can be more
then $2/11$ on Fig.7 (white inner figures)
 for the case of both hierarchies. We do it only for the mass $M=1$
 GeV because this regions are at small values of $\varepsilon$ and it will not intersect with other subsequent
 constrains. Moreover, at some values of phases
 leptonic asymmetry in this region can be less then $2/11$ and so we can not
 exclude this region ultimately.

By way of example, we present possible values of $I$ sterile
neutrino decay rate $\Gamma_I$ \eqref{totalrate} and rate of
oscillations between $I$ and $J$ sterile neutrinos $\Gamma_{IJ}$
\eqref{IJ} for $M=1$ GeV and $\theta_{13}=10^{\rm o}$ on Fig.8. As
one can see the values of $\Gamma_I$, $\Gamma_{IJ}$ are really of
the same order as $\Delta M$. It confirms previous assumption about
necessity of taking into consideration oscillations between sterile
neutrinos.

\begin{figure}[t]
\qquad\qquad\qquad\qquad\includegraphics[width=80mm]{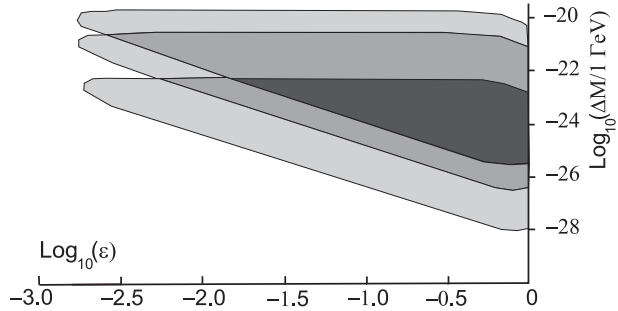}
\caption{The grey areas are the regions of parameters where $\Delta
> 10^{-3}$ for the case of the normal hierarchy. The areas correspond to $M = 0.3$ GeV (bottom), $M = 1$
GeV (middle), $M = 2$ GeV (top).}
\end{figure}

\begin{figure}[b]
\includegraphics[width=162mm]{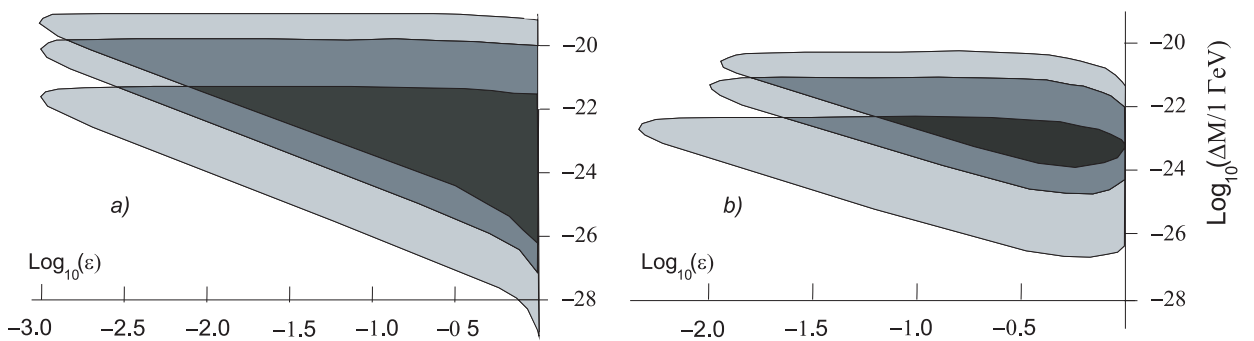}
\caption{The grey areas are the regions of parameters where $\Delta
> 10^{-3}$ for the case of the inverted hierarchy. The areas correspond to $M = 0.3$ GeV (bottom), $M = 1$
GeV (middle), $M = 2$ GeV (top). Figures \textit{a}) and \textit{b})
represent the case of $\theta_{13}=0^{\rm o}$ and $10^{\rm o}$
correspondingly.}
\end{figure}

In order to create a leptonic asymmetry the sterile neutrinos should
be out of thermal equilibrium.  That means that
\begin{equation}\label{out}
\Gamma_2\lesssim H,
\end{equation}
where $H$ is Hubble parameter that determines the expansion rate of
the Universe. In the radiative dominated epoch Hubble parameter is
given by
\begin{equation}\label{Hubble}
H=\frac{T^2}{M_{PL}^\ast}
\end{equation}
where $M_{PL}^\ast=\sqrt{\frac{90}{8\pi^3g^\ast(T)}}M_{PL}$,
$M_{PL}=1.22\cdot10^{19}$ GeV is the Planck mass, $g^\ast(T)$ is the
internal degrees of freedom \cite{kolb}. At temperature $T\sim1$ GeV
we can take $g^*\simeq65$.

So we get condition
\begin{equation}
\sqrt{M_{PL}^*\Gamma_2}\lesssim T.
\end{equation}

The out-of-equilibrium condition means that sterile neutrinos should
decay at a temperature smaller than their mass ($T\lesssim M$).
Moreover the sterile neutrinos should decay before the creation of
DM so that the leptonic asymmetry enhances the DM production. The DM
is created at $T\sim 0.1$ GeV. Therefore,
\begin{equation}\label{dmcon}
 0.1\lesssim\frac{\sqrt{M_{PL^*}\Gamma_2}}{1{\rm
GeV}}\lesssim \frac{M} {1\rm GeV}.
\end{equation}

\begin{figure}[t]
\includegraphics[width=162mm]{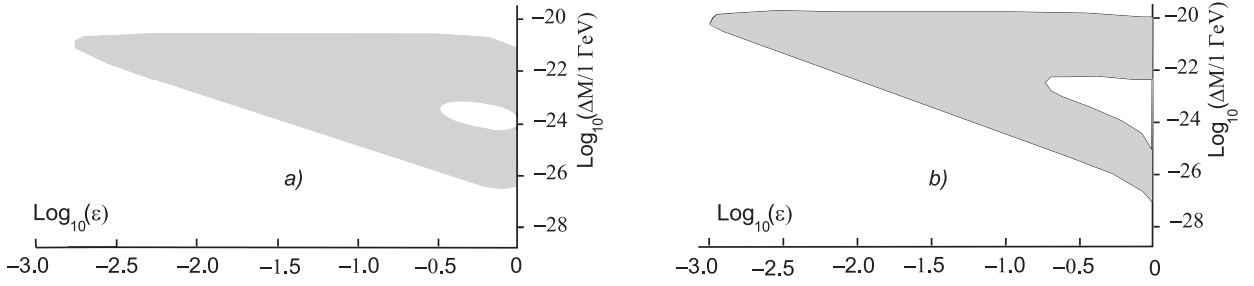}
\caption{The grey areas represent regions of parameters where
$10^{-3}<\Delta<2/11$ for $M = 1$ GeV in case of normal (\textit{a})
and inverted hierarchy (\textit{b}).}
\end{figure}

\begin{figure}[b]
\includegraphics[width=155mm]{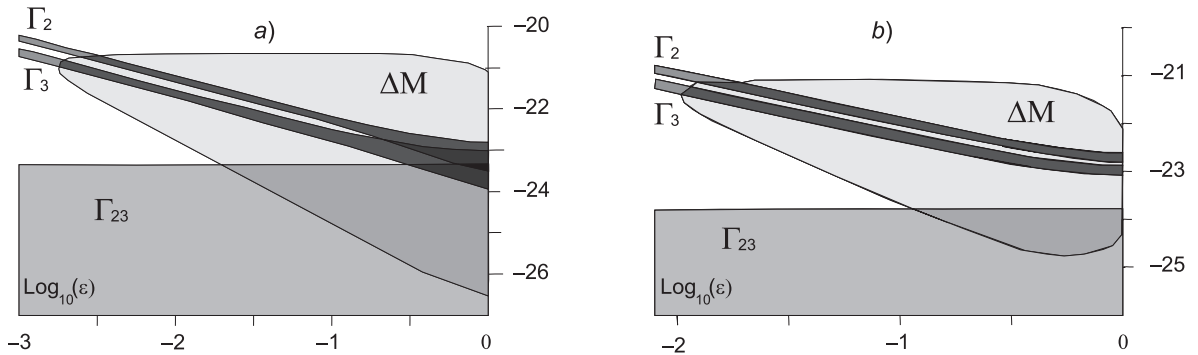}
\caption{The values of rates $Log_{10}(\Gamma_I/1$GeV),
$Log_{10}(\Gamma_{23}/1$GeV) and $Log_{10}(\Delta M/1$GeV) for
leptonic asymmetry $\Delta>10^{-3}$ for the case of $M=1$GeV and
$\theta_{13}=10^{\rm o}$ are on the ordinate axis: \textit{a}) the
case of normal hierarchy (A type of solution), \textit{b}) the case
of inverted hierarchy (B type of solution).}
\end{figure}

\begin{figure}[t]
\qquad\qquad\qquad\qquad\qquad\includegraphics[width=85mm]{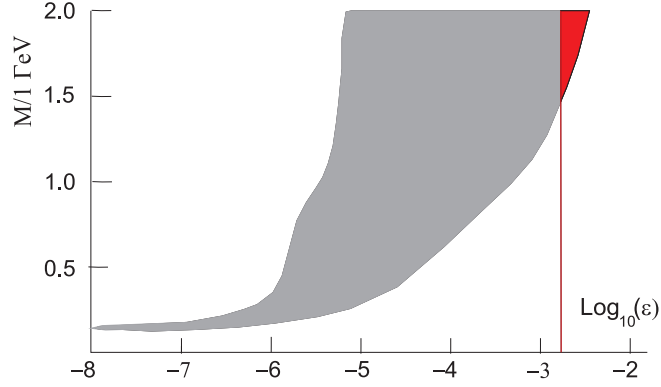}
\caption{The case of the normal hierarchy. The points on grey and
red regions satisfy constraint \eqref{dmcon}. The region on the
right from vertical red line satisfies condition \eqref{asymmetry}
also.} 
\end{figure}

We illustrate  on Fig.9 the region of values of $M$ and
$\varepsilon$ where condition \eqref{dmcon} is satisfied for the
case of the normal hierarchy. At scale of parameters presented on
Fig.9 the difference between the case of $\theta_{13}=0$ or
$\theta_{13}=10^{\rm o}$ and between the case of solution of A or B
types is small, so we present only one figure. It is not true for
the case of the inverted hierarchy, see Fig.10.

It should be noted that region on Fig.9 is almost not sensitive to
value of $\theta_{13}$ at range $0<\theta_{13}<16^{\rm o}$. In case
of the inverted hierarchy it is true for the regions on Fig.10
\textit{b}) and $\theta_{13}<18^{\rm o}$, but for the case of
$\theta_{13}=0$ the regions are appreciably different.

As one can see there are regions on Fig.9 (red) and Fig.10
\textit{a}) (red and blue) where conditions \eqref{asymmetry} and
\eqref{dmcon} are satisfied simultaneously. This region of
parameters is suitable for DM production in the $\nu MSM$. For the
case of inverted hierarchy and nonzero value of $\theta_{13}$ we
have no region that is suitable for DM production. So, in $\nu$ MSM
for physical nonzero value of $\theta_{13}$ and mass of sterile
neutrino $m_\pi<M<2$ GeV DM production can be realized only in case
of normal hierarchy of active neutrino mass.

The region suitable for DM production (the case of the normal
hierarchy and nonzero $\theta_{13}$) can be used to obtain
constraints for mass splitting of the sterile neutrino. Fixing mass
of the sterile neutrino one obtains possible values of $\varepsilon$
(see Fig.9) and using Fig.5 one can obtain possible values of the
mass splitting for the sterile neutrino with mass $M$. If mass of
the sterile neutrino is on lower boundary of the allowed mass range
($M\simeq 1.4$ GeV) than value of $\Delta M$ is exactly known
($\Delta M\approx 5\cdot10^{-21}$ GeV). If mass of the sterile
neutrino is on upper bound of the allowed mass range ($M=2$ GeV)
than $\Delta M$ can possess the values from the range
$10^{-21}\lesssim\Delta M/1 {\rm GeV}\lesssim10^{-20}$.

\begin{figure}[t]
\includegraphics[width=160mm]{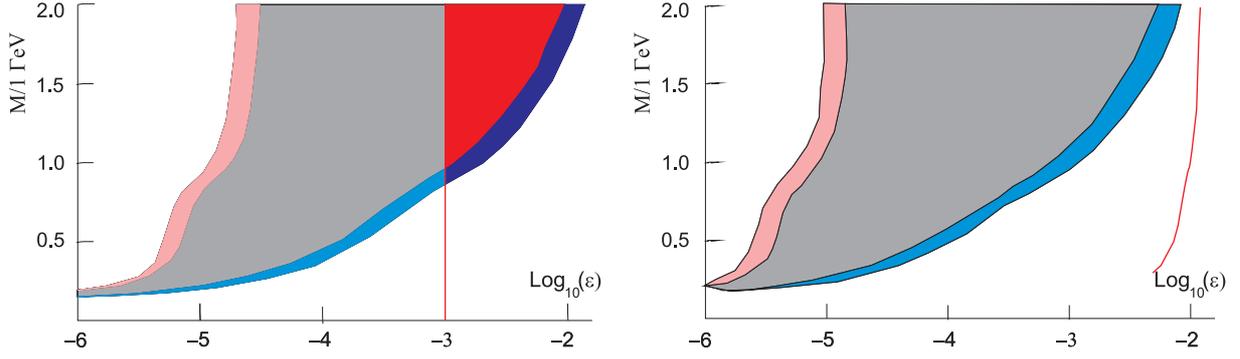}
\caption{The case of the inverted hierarchy: a) $\theta_{13}=0$,
\textit{b}) $\theta_{13}=10^{\rm o}$. The pink, grey and red regions
corresponds to the A type of solutions. The grey, red, sky blue and
blue regions corresponds to the B type of solutions. The points on
this regions satisfy constraint \eqref{dmcon}. The region on the
right from red line satisfies condition \eqref{asymmetry} also.} 
\end{figure}

Some existing experimental data restrict the area of parameters of
$\nu MSM$. For $M < 0.45$ GeV the best constraints come from the
CERN PS191 experiment. For $0.45 < M < 2$ GeV the constraints come
from the NuTeV, CHARM and BEBC experiments. The range of parameters
admitted by these experimental data is summarized in \cite{ShProg}.
These parameters are the mixing angle $(\Theta^+\Theta)_{22}$ (it
defines the range of reactions with sterile neutrino)  and the mass
of the heavier sterile neutrino $M$.

To compare obtained  in the present paper constraints on the $\nu
MSM$ parameters (see Fig.9 and Fig.10) with constraints summarized
in \cite{ShProg} one has to rebuild allowed regions in the space of
parameters $M$ and $\theta^2_{\nu N_2}=(\Theta^+\Theta)_{22}$.

In general case the relation between $(\Theta^+\Theta)_{22}$ and
$\varepsilon$ is quite difficult. Really, in accordance with
\eqref{transition}
 and\eqref{theta} we have
\begin{equation}\label{Theta22}
 (\Theta^+\Theta)_{22}=\frac{\nu^2}{2M^2} (V^+h^+hV)_{22}=\frac{\nu^2}{2M^2}(F_2^2+F_3^2-2
|h^+h|_{23}\sin\chi),
\end{equation}
where $\chi=arg[(h^+h)_{23}]$. Using  \eqref{m2m3} --
\eqref{epsilondef} we get
\begin{equation}\label{F2F3}
 F_2^2=\frac{M}{\nu^2\varepsilon}(m_c+m_b),\quad F_3=\varepsilon
F_2,\quad |h^+h|_{23}=\frac{M}{\nu^2}(m_c-m_b)
\end{equation}
and
\begin{equation}\label{Theta22F}
 (\Theta^+\Theta)_{22}=\frac{m_c+m_b}{2M\varepsilon
}\left(1+\varepsilon^2-2\frac{m_c-m_b}{m_c+m_b}\varepsilon
\sin\chi\right).
\end{equation}
The problem is in parameter $\chi$ that is a complicated function of
many parameters. But for our case (see Fig.9 and Fig.10,
$\epsilon<0.16$) we can use approximate relation
\begin{equation}\label{Theta22Final}
 (\Theta^+\Theta)_{22}=\frac{m_c+m_b}{2M\varepsilon}.
\end{equation}

The imposition of our constraints are presented on Fig.9 and Fig.10
for nonzero value of $\theta_{13}$ and summarized constraints from
\cite{ShProg} is presented 
 on Fig.11. Above the line marked
"BAU", baryogenesis is not possible: here sterile neutrinos
 come to thermal equilibrium above the $T_{EW}$ temperature. Below the line marked "See-saw", the data
on neutrino masses and mixing cannot be explained using "see-saw"\,
mechanism. The region noted as "BBN" is disfavoured by the
considerations of Big Bang Nucleosynthesis.  The region marked
''Experiment'' shows the part of the parameter space excluded by
direct searches for singlet fermions. The regions market "Cos"\,,
"$\Delta$"\, and "DM" were builded in this paper. The grey and blue
region "Cos"\,  shows the parametric space allowed by cosmological
constraint \eqref{dmcon} (grey region corresponds to A and B type of
solution, blue region  corresponds to  B type of solution), the
dashed region market "$\Delta$"\, shows the parametric space allowed
by constraint \eqref{asymmetry}, the red region marked "DM"\, shows
the parametric space where constraints \eqref{asymmetry} and
\eqref{dmcon} are noncontradictory. The last region is preferred for
DM production according to calculations of the present paper.

\begin{figure}[t]
\includegraphics[width=165mm]{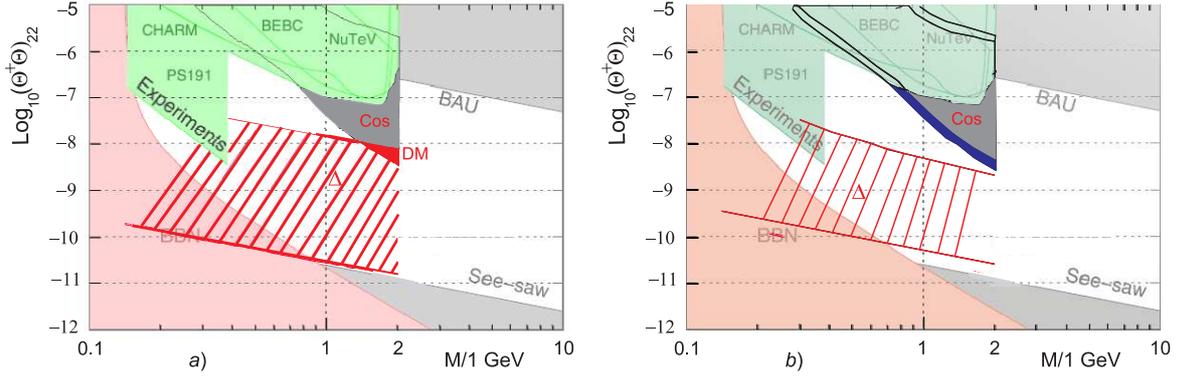} \caption{The imposition
of our constraints and summarized constraints from \cite{ShProg}:
\textit{a}) the case of the normal hierarchy, \textit{b}) the case
of the inverted hierarchy.}
\end{figure}

The red region marked "DM"\, is shown on Fig.12  in the scaled-up
form.  The difference between the case of $\theta_{13}=0^{\rm 0}$ or
$\theta_{13}=10^{\rm o}$, and between type of A or B solutions is
illustrated. As one see the choice of solutions of A or B type makes
greater change in the allowed region then the choice of
$\theta_{13}=0^{\rm 0}$ or $\theta_{13}=10^{\rm 0}$.

\section{Conclusion}\label{rozdil5}

In the present paper we consider the leptonic asymmetry generation
at $T\ll T_{EW}$ when the masses of  two heavier sterile neutrinos
are between $m_\pi$ and 2 GeV.

We conclude that oscillations and decays of sterile neutrinos can
produce a leptonic asymmetry that is large enough to enhance the DM
production sufficiently to explain the observed DM in the Universe,
but only for the case of the normal hierarchy of the active neutrino
mass. The allowed range of parameters is narrow and it is presented
on Fig.11 and Fig.12. It should be noted that allowed mass range for
heavier sterile neutrino is $1.42 (1.55)\lesssim M<2$ GeV for B (A)
type of solutions and the mixing angle between active and sterile
neutrino is $-7.91(-7.98)\lesssim{\rm
Log}_{10}(\Theta^+\Theta)_{22}\lesssim-8.41(-8.35)$ for B (A) type
of solutions. If mass of the sterile neutrino is on lower boundary
of the allowed mass range than value of $\Delta M$ is exactly known
($\Delta M\approx 5\cdot10^{-21}$ GeV). If mass of the sterile
neutrino is on upper bound of the allowed mass range  than $\Delta
M$ can possess the values from the range $10^{-21}\lesssim\Delta M/1
{\rm GeV}\lesssim10^{-20}$. For the case of the inverted hierarchy
there is no region suitable for DM production.

The big range of parameters of the $\nu MSM$ is not forbidden by the
existing experimental data, see Fig.11.  Combining of this range
with our constraints (red region "DM"\, on Fig.11) leads to
conclusion that improvement of previous experiments, as NuTeV or
CHARM, of one or two order of magnitude can  exclude the $\nu MSM$
with $M < 2$ GeV or detect the right-handed neutrinos.

It should be noted that our constraints are quite a rough and can be
used only for estimation. Really, the form of red region "DM"\, is
very sensitive to cosmological constraints. Applied condition $ 0.1
{\rm GeV} <\sqrt{M_{PL^*}\Gamma_2}<M $ is very approximate. The
correct description of the processes can be made in frame of the
density matrix formalism or Boltzmann equations. Our computation is
not valid for $M > 2$ GeV. However, the extrapolation of our result,
see Fig.11,  suggests that the range of admitted parameters for the
case of the normal hierarchy becomes bigger for masses above 2 GeV.
We expect that for masses above 2 GeV DM production can be realized
for the case of inverted hierarchy too.

\begin{figure}[t]
\qquad\qquad\quad\includegraphics[width=115mm]{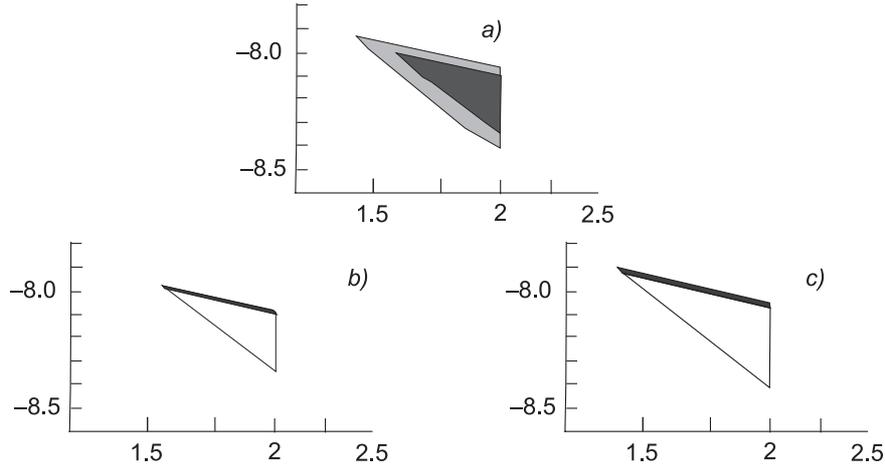} \caption{The
red region "DM" from Fig.10  in the scaled-up form: \textit{a})
$\theta_{13}=0^{\rm 0}$ type A (dark) and type B (light);
\textit{b}) type A: $\theta_{13}=0^{\rm 0}$  (white) and
$\theta_{13}=10^{\rm 0}$ (black); \textit{c}) type B:
$\theta_{13}=0^{\rm 0}$ (white) and $\theta_{13}=10^{\rm 0}$
(black). The variable M/1 GeV is along the abscissa axis and the
variable ${\rm Log}_{10}(\Theta^+\Theta)_{22}$ is along the ordinate
axis.}
\end{figure}

During computation we used two types (A or B) of solutions
\eqref{solution}. This is due to the fact that ratios of the Yukawa
matrix elements (enter into the expression for the mixing angle
$\Theta_{\alpha I}$) can be expressed through solutions
\eqref{solution} with two possible choice of sign consistent with
condition \eqref{a10}. It is closely related to the symmetry of
\eqref{system} under replacing the elements of the second column of
the Yukawa matrix by elements of the third column.  This two
variants are equal in rights.

The computation of the leptonic asymmetry in the applied simple
model allows us to make some conclusions that, seemingly, will be
correct and under more rigorous consideration. Namely, the initial
state of the right-handed neutrino in form \eqref{roini} are not
important for lepton asymmetry generation (the final results are not
sensitive to values of the constants $\alpha_I$). For the case of
normal hierarchy the deviation of the mixing angle $\theta_{13}$
from its zero value (up to value $16^{\rm 0}$) almost does not
change the region suitable for DM production.  For the case of
inverted hierarchy results are different for $\theta_{13}=0$ and
$\theta_{13}\neq0$. Our calculations indicates that case of
$\theta_{13}=0$ leads to existing of region suitable for DM
production, but at nonzero values of $\theta_{13}$ this region does
not exist. Values of $\theta_{13}$ in range $\theta_{13}<18^{\rm o}$
($\theta_{13}\neq0$) almost does not change the region suitable for
DM production.

It's essential to note that during computations we have used
functions maximized over unknown parameters of the model (phases
$\delta,\xi,\alpha_{2},\alpha_{1}$). If the maximization procedure
was not performed the final functions are  sensitive to values of
mentioned phases. So, the obtained results are very optimistic. But
if the proposed on Fig.11 region of parameters "DM"
will be forbidden
by experiment data it will mean that mass of heavier sterile
neutrinos must be lager 2 Gev.

An essential assumption we have made is that the background effects
are negligible.  We do not have justify that it can be neglected in
the thermal bath of the universe. For simplicity the computations
were made at zero temperature. A rigorous justification of this
assumptions is needed.

It should be noted that region suitable for DM production in $\nu
MSM$ was recently calculated in frame of more general formalism in
\cite{Drewes}. Certainly,  results of \cite{Drewes} somewhat differ
from our simple calculations.

\section*{Acknowledgments}

We would like to thank Marco Drewes and Tibor Frossard  for the idea
of treating this subject, and for useful comments and discussions.
This work has been supported by the Swiss Science Foundation (grant
SCOPES 2010-2012, No. IZ73Z0\_128040).

\bibliographystyle{unsrt}

\end{document}